\documentclass [pra,noeprint,reprint,groupedaddress,showpacs]{revtex4-1}
\usepackage{graphicx}
\usepackage{dcolumn}
\usepackage{amsmath}
\usepackage{amsfonts}
\usepackage{amssymb}
\usepackage{bm}
\usepackage{nicefrac}
\usepackage{color}
\usepackage{float}
\usepackage{sidecap}
\graphicspath{{figures/}}
\usepackage{hyperref}
\usepackage[none]{hyphenat}
\hypersetup{backref,
pdfstartview=Fit,
colorlinks=true,
citecolor=blue,
urlcolor = blue,
linkcolor = blue}

\begin{document}
\title{Continuous microwave hole burning and population oscillations in a diamond spin ensemble}

\author{Haitham A.R. El-Ella}
\email{haitham.el@gmail.com}
\author{Alexander Huck}
\author{Ulrik L. Andersen} 
\affiliation{Centre for Macroscopic Quantum States (bigQ), Department of Physics, Technical University of Denmark, 2800 Kongens Lyngby, Denmark}

	\begin{abstract}
	Continuous spectral hole burning and spin-level population oscillations are studied in an inhomogeneously broadened diamond-based spin ensemble composed of substitutional nitrogen and nitrogen-vacancy centres created through neutron irradiation and annealing. The burnt spectral features highlight a detuning-dependent homogeneous hole linewidth that is up to three orders of magnitude narrower than the total inhomogeneous ensemble linewidth. Continuous population oscillations are observed to quickly decay beyond a pump and probe detuning of 5 Hz, and are numerically modelled using a five-level system of coupled rate equations. Fourier analysis of these oscillations highlight discrete $^{13}$C hyperfine interactions, with energies within the inhomogeneous ensemble linewidth, as well as suspected nuclear $3/2$-spin coupled signatures likely related to the $^7$Li byproduct of neutron irradiation.  
	\end{abstract}
\maketitle
	
	\section{Introduction}
	
	Spectral hole burning is a conventional spectroscopic technique first employed in nuclear magnetic resonance measurements for investigating the broadening mechanisms underlying non-Lorentzian spectral features \cite{Moerner1988}. This technique suppresses or ``burns'' part of the broadened spectrum by saturating the emitters at a given frequency, creating discontinuities with linewidths that can be orders of magnitude narrower than the regular spectral features. Due to the
	hole’s extreme sensitivity to perturbations of both radiative and nonradiative ensemble transitions, burnt holes convey unique information on the ensemble's coupled frequencies and relaxation dynamics, complimentary to that obtained through pulsed double-resonance schemes. Here we report on continuous hole burning experiments and the analysis of the resulting population oscillations (POs) in a spin ensemble consisting of substitutional nitrogen (P1) spins and a nitrogen vacancy (N-$V^-$) spin subensemble, in a neutron irradiated and annealed high-pressure-high-temperature grown diamond.	

	The N-$V^-$ system is unprecedented in its use for quantum-limited metrology and is continuously pushing boundaries on electron- and nuclear-based magnetic sensing \cite{Lovchinsky2017,Glenn2018}. While considerable work has been carried out in exploring the noise and broadening dynamics of N-$V^-$s using a variety of schemes \cite{DeLange2012,Dreau2014,Bar-Gill2012,Mittiga2018,Rosenzweig2018,Bauch2018}, there is still value in further investigating broadening under continuous-wave (CW) schemes, and the distinct properties of adiabatic continuous POs. This is motivated by the desire to understand and use N-$V^-$  ensembles that are dense enough to demonstrate physical novelty and superior technical applicability \cite{Taylor2008, Breeze2018}, while being limited by the difficulty of creating homogeneous and pure ensembles \cite{Acosta2009,Chen2018} and consequently, their short coherence times ($T_2{<}100$ ns), which renders pulsed-based experiments challenging.
	
	The basic experimental premises lies in applying two driving fields; a pump field is continuously applied at a variable frequency within the broadened spectrum to saturate a portion of the emitters, such that when a second probe field is swept across the pump field frequency, a sharp dip in amplitude (a ``burnt'' hole) is detected. In a pulsed scheme, the generated hole width is inversely proportional to the pump duration and an indefinitely extended pulse time will generate a linewidth proportionally limited by the coherence time of the resonant subensemble \cite{Moerner1988}. The same basic configuration is a staple tool in experimental atomic and quantum optics when varying the group velocity of light pulses propagating through a saturable absorptive medium, e.g. \cite{Phillips2001}, or implementing optical memories, e.g. \cite{Maynard2014}. This technique has also been previously employed in the study of N-$V^-$ ensembles, where it has been shown to suppress broadening within a selected bandwidth and enhance thermometery \cite{Kehayias2014}, to demonstrate optically detectable POs of the N-$V^-$ magnetic ground-state spin transitions \cite{Mrozek2016}, and to demonstrate long-lived dark-states of hybrid cavity-coupled N-$V^-$ spin ensembles to be used as optical memories \cite{Putz2017}. Here we experimentally demonstrate the spectral limit of continuous hole burning and the features and utility of the resulting POs in measuring and identifying nuclear spins coupled to the N-$V^-$ ensemble.
	
	\begin{figure*}
		\centering
		\includegraphics[scale = 0.8]{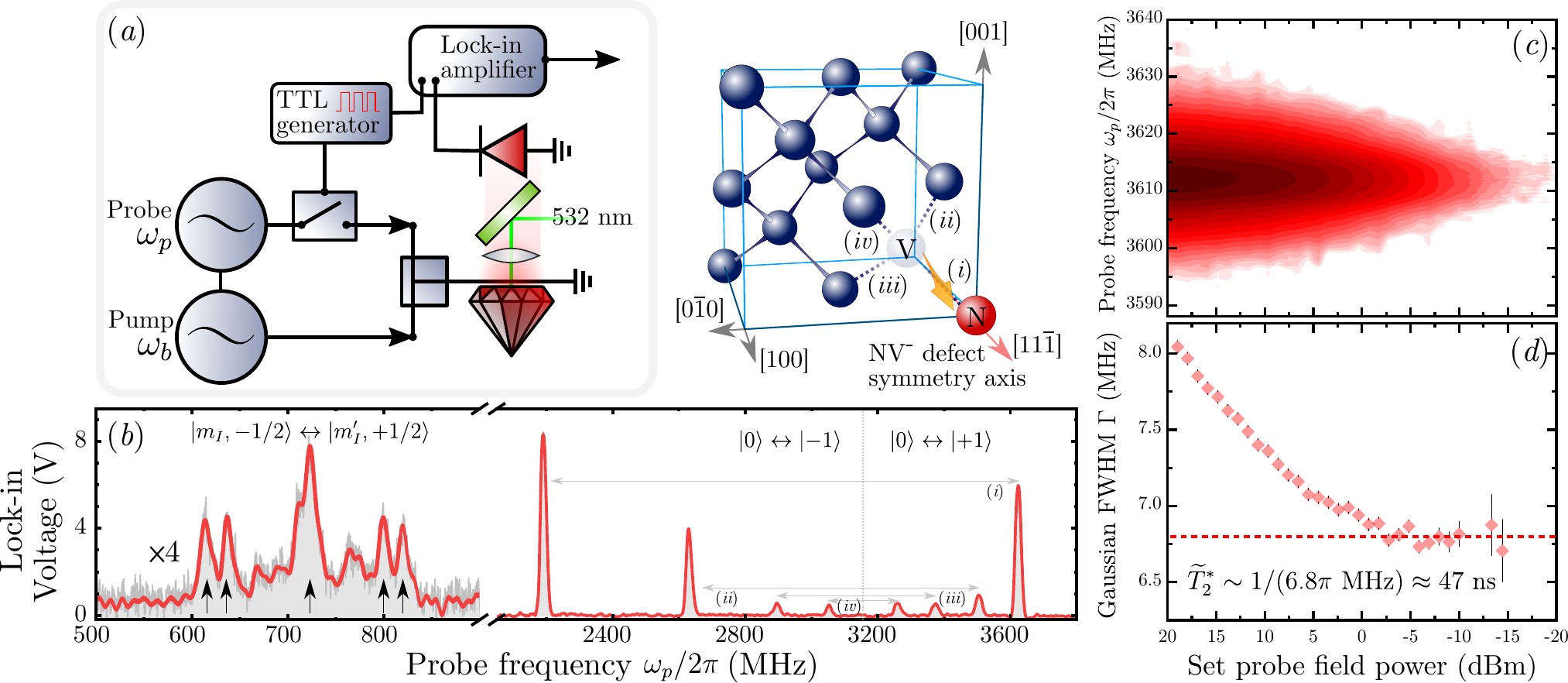}
		\caption{(a) Summary of the experimental setup used. (b) Typical ODMR measured from the diamond spin ensemble under a $\sim30$ mT field, also showing the coupled P1 spin resonances. The four crystallographic distinct N-$V^-$ sub-ensembles are annotated with reference to the unit cell schematic. (c) ODMR map of the highest energy $\vert0\rangle{\leftrightarrow}\vert{+}1\rangle$ spin resonance as a function of the probe field power, highlighting the unresolvable hyperfine splitting as the power is decreased due to inhomogeneous broadening. Note that a logarithmic colour scale. (d) The overall linewidth plateaus as the power is decreased, to a near constant linewidth which is proportional to the collective pure dephasing rate $\widetilde{T}_2^*$, in this case being in the order of 50 ns.}		
	\end{figure*}   
	
	The inhomogeneity of a spin ensemble is caused by a spatially variant electric and magnetic environment which may both shift the transition frequencies and degrade the coherence times in portions of an ensemble. Such a varying electromagnetic environment may be a consequence of extrinsic field variations with respect to the probed ensemble volume, or it may be intrinsic due to the presence of spatially fluctuating spin impurities that couple to portions of an ensemble via, e.g., Fermi-contact and/or dipole-mediated mechanisms. The resulting inhomogeneity is therefore usually caused by an accumulation of intrinsic and extrinsic perturbations, and its characterization is dependent on the probed ensemble volume.
	
	Solid crystal-based spin ensembles that are coupled to environmental spins also generate discrete magnetic satellite resonances due to discrete distance-dependent magnetic coupling. These coupled states may be measured using either CW or pulsed double-resonance schemes, however, resolving and identifying coupled frequencies which lie within an inhomogeneous linewidth is not possible in a direct CW measurement, and nontrivial using pulsed schemes. Such resonances may alternatively be detected by generating and analysing POs; the procedure of adiabatically cycling the population of a subset of an ensemble also cycles the population of spins coupled to this subset. This drives their coupling frequency, which is superimposed with the POs frequency and will therefore be evident in the POs Fourier spectrum.
	
	In the experiments reported here, a lock-in scheme is employed for CW optical detection of magnetic resonance (ODMR) while a second pump field is applied simultaneously for burning spectral holes, analogous to that demonstrated in Ref. \cite{Kehayias2014}, but instead of $\pi$-pulsing the pump field, the probe field is modulated for lock-in detection under continuous laser excitation. Using this configuration, generated hole widths (full width at half maximum) highlight a detuning-dependent homogeneous component of the ensemble, which outlines the ensemble’s dominant broadening mechanism. The generation and measurement of POs is also presented, and its characteristics are modelled and discussed using a typical  N-$V^-$ five-level system of coupled rate equations. Finally, the Fourier components of the POs are mapped out, from which both dressed states and hyperfine coupled states, in particular those associated with $^{13}$C nuclear spins, are identified at energies well within the ensemble’s inhomogeneous linewidth.
					
	\section{Experiment \& Spin Ensemble}
	
	A confocal experimental arrangement is employed, as summarized in Fig. 1(a), using an objective condenser lens (NA $\approx0.8$) with a focal spot diameter in the order of ${\sim}2$ $\mu$m. A 30-mW, 532-nm laser is continuously focused onto the diamond surface, and the resulting N-$V^-$ ensemble fluorescence is collected using the same objective and passed through a dichroic mirror (cutoff at 540 nm) and two long pass filters (cutoff at 600 nm), before hitting a 1 k$\Omega$ biased Si photo-diode (Thorlabs DET36A, $\sim$4 MHz bandwidth). The delivery of two microwave fields was carried out using a single $20~\mu\text{m}$-diameter copper wire which was tightly pulled across the surface of the diamond. Phase-locked pump and probe microwave fields, with respective angular frequencies $\omega_b$ and $\omega_p$, were generated using the same source (Windfreak Technologies, dual channel SynthHD). For measuring an ODMR spectrum, the microwave probe field amplitude is digitally modulated at 10 kHz, and detection is locked in at the resulting fluorescence modulation (using a Stanford Research Systems SR510), under continuous laser illumination.
	
	The diamond used here is a polished 1b diamond crystal (Element Six), with approximate dimensions of $0.5~\text{mm}\times9~\text{mm}^2$, a natural abundance of $^{13}$C isotopes ( $\sim1$ \%), and a specified P1 ($^{14}$N in particular) concentration in the order of 200 ppm, which was subjected to thermal neutron irradiation and annealing at 900 ◦C. A trace amount of $^{60}$Co and $^{7}$Be  atoms are implanted during neutron irradiation, which was confirmed to be present using dosimetry. These isotopes radioactively decay into $^{60}$Ni and $^{7}$Li, respectively, the latter possessing a 3/2 nuclear spin. A minimum, lower-bound N-V − concentration was estimated using confocal microscopy to be in the order of 0.5 ppm. However, due to the dominant P1 concentration, significant quenching of the N-$V^-$ fluorescence occurs \cite{GattoMonticone2013}, and the actual concentration is expected to be at least in the order of 10 ppm.
	
	A model of the defect arrangement within the diamond unit cell, and a typical lock-in amplified ODMR spectrum is shown in Fig. 1(b), which was measured with an applied magnetic field at approximately $\sim30$ mT, and a $\sim20^\circ$ angle to all $\langle111\rangle$ crystallographic orientations. The resulting spectrum highlights Zeeman-split spins transitions $\vert0\rangle{\leftrightarrow}\vert{\pm}1\rangle$ associated with the four crystallographically distinct subensembles (annotated in the unit cell diagram), as well as the resonance frequencies of the P1 electron spin ensemble. The visible P1 electron spin resonances occur here between 600 and 850 MHz, and consist of nine peaks based on the hyperfine coupling of the P1 electron ($S{=}1/2$) and quadrupole nuclear ($I{=}1$) spins: five peaks originating from allowed $\vert m_I, {-} 1/2\rangle{\leftrightarrow}\vert m_I', 1/2\rangle$ transitions, and four low amplitude peaks related to nuclear spin flip-flop and forbidden transitions ($\Delta m_I \neq\{0, 1\}$) \cite{Hall2016}. These normally “dark” P1 resonances are directly detectable in this crystal through the N-$V^-$ fluorescence, in this case likely due to a Raman-based mechanism of cross-relaxation between the N-V − and P1 electron spins, as discussed in Ref.\cite{Purser2019}. In any case, this reflects their dominant presence and therefore as the most likely dominant cause of the observed inhomogeneous broadening.
	
	A direct estimate of the ensembles collective pure dephasing time $\widetilde{T}_2^*$ may be obtained by plotting the spectral linewidth as a function of applied microwave power, as shown in Figs. 1(c) and 1(d). The measured linewidth is reciprocally proportional to the ensembles collective dephasing rate, $\Gamma = 1/\pi \widetilde{T}_2$, for which $1/\widetilde{T}_2$ is a sum of the collective spin lifetimes $\widetilde{T}_1$ and $\widetilde{T}_2^*$, the latter being an accumulation of all the intrinsic and extrinsic sources of decoherence affecting the ensemble, such as that induced by the nuclear spin bath or microwave power broadening. By ensuring that the measurement is performed within a regime where the optical power and detected fluorescence are linearly dependent, and assuming that $\widetilde{T}_1>>\widetilde{T}_2^*$, an estimate of $\widetilde{T}_2^*$ may be obtained. As the microwave power is decreased, power broadening is reduced, and plateaus to a near constant level around 6.8 MHz, representing a point where T$\widetilde{T}_2 \sim\widetilde{T}_2^*\approx 50$ ns. The severity of the broadening prevents the measurement of the hyperfine transitions resulting from coupling between the N-$V^-$ electron spin and its $^{14}$N nuclear spin, which should outline three overlapping peaks separated by 2.16 MHz. This is evident in Fig. 1(c) where the resonance peak becomes unresolvable before the presence of three hyperfine peaks becomes resolvable, or even before the lineshape shifts from being a single Gaussian to a measurable non-Gaussian sum of three peaks. It is worth mentioning that the estimated $\sim\widetilde{T}_2^*\approx 50$ ns for a P1 concentration in the order of 200 ppm is consistent with the reported P1 concentration-dependent trend of $\sim\widetilde{T}_2^*$ in \cite{Bauch2019}.
	
	\section{Continuous Hole Burning}
	
	\begin{figure*}
		\centering
		\includegraphics[scale = 0.33]{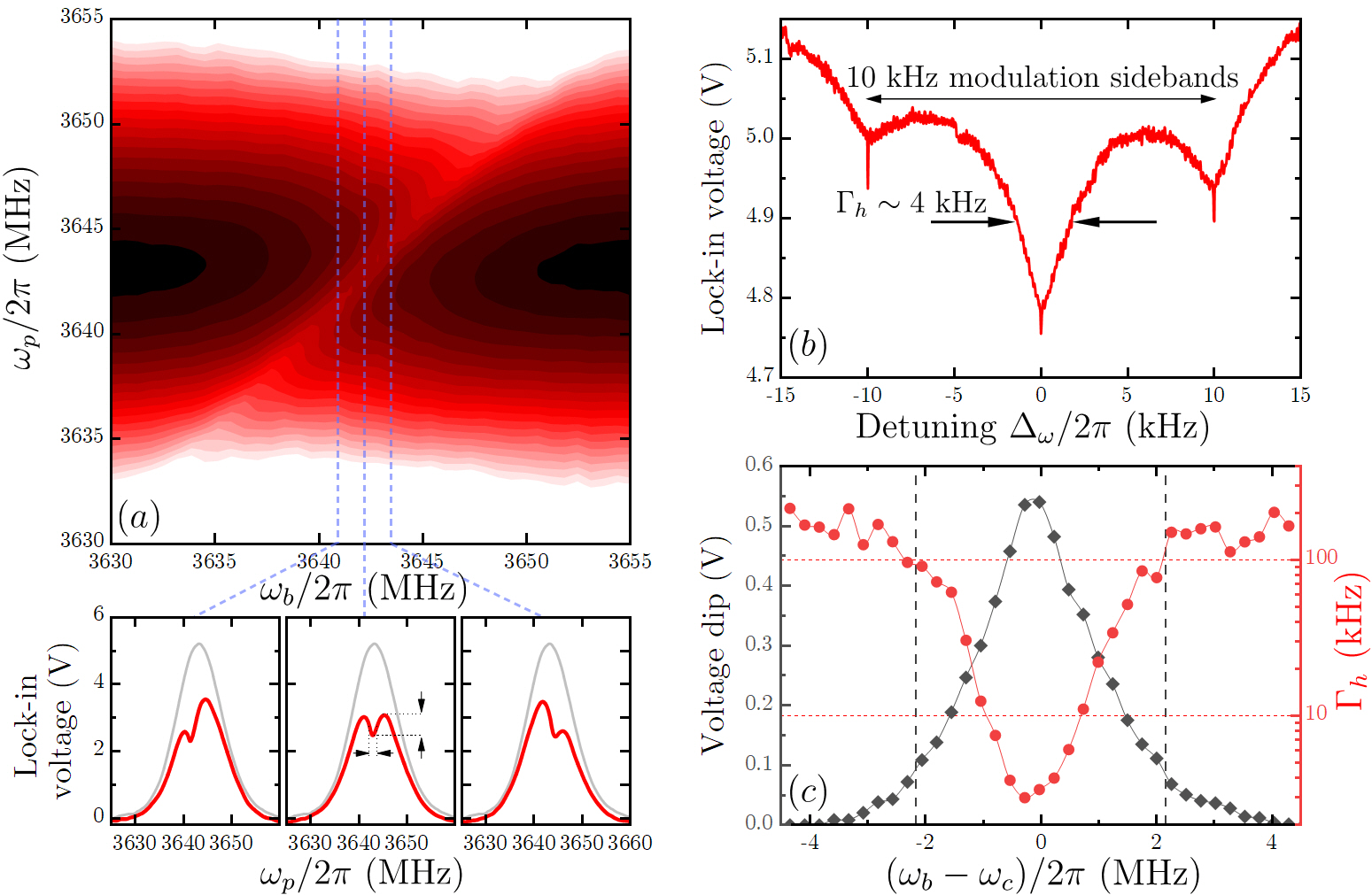}
		\caption{(a) Measured ODMR map using a logarithmic colour scale, as a function of the pump and probe field frequencies. (b) Zoom-in on the burnt hole spectrum where $\Delta_\omega=\omega_p-\omega_b$, highlighting the 10 kHz modulation side-bands of the probe field $\omega_p$. (c) Plot of the extracted voltage drop and hole linewidth $\Gamma_h$, as annotated in the central subplot of (a), as a function of detuning between the pump field and the central frequency of the spin transition $\omega_c$.}
	\end{figure*}  
	
	A typical hole-burnt spectral map is shown in Fig. 2(a). Under CW-ODMR, the application of a microwave field resonant with a spin transition during continuous laser illumination results in a drop of the N-$V^-$ fluorescence, due to the spin-dependent shelving dynamics of its level structure \cite{Doherty2013}. The application of a modulated microwave field under the same scheme results in a modulation of the fluorescence, enabling lock-in amplification. When both the continuous and modulated fields are simultaneously applied, the resulting modulated fluorescence amplitude is decreased. This is caused by a saturation of a portion of the ensemble, which becomes effectively shielded from the modulated driving field, thereby resulting in a reduced lock-in voltage.
 
 	An example of a spectrum with the narrowest hole linewidth $\Gamma_h$ is plotted in Fig. 2(b) as a function of the detuning between the pump and probe field frequencies $(\Delta_\omega = \omega_p-\omega_b)$, as well as the detuning-dependent characteristics of the pump and the central spin frequencies $(\omega_b-\omega_c)$ in Fig. 2(c). The detuning dependence of $\Gamma_h$ is measured by fitting a Lorentz function, while the signal drop is measured by
 	direct extraction of the difference between the lowest shoulder and the burnt hole amplitude as annotated in the extracted plot of Fig. 2(a). The hole amplitude relative to the total peak amplitude is directly dependent on the pump field power, while the hole’s full-width-at-half-maximum $\Gamma_h$ is somewhat insensitive as long as the optical excitation rate exceeds the microwave excitation rate to avoid any power-broadening (ensuring the so-called linewidth-narrowing regime \cite{Jensen2013}). The measured spectrum and associated hole linewidths are well
 	represented by the example plotted in Fig. 2(b), which shows two side bands due to the 10 kHz modulation of the probe
 	field, and a central dip with $\Gamma_h\sim4$ kHz. In general, $\Gamma_h$ is
 	equal to twice the homogeneous linewidth of the ensembles
 	homogeneous component \cite{Moerner1988}:
	\begin{align}
	2\Gamma_h =(\pi T_2)^{-1} = (2\pi T_1)^{-1} + (\pi T_2^*)^{-1}. 
	\end{align}
	As opposed to the ensemble $\widetilde{T}_2^*$, $\Gamma_h$ highlights a detuning dependent homogeneous coherence time with the range $T_2\sim$ 40 $\mu$s - 0.7 $\mu$s. While the voltage drop naturally outlines the detuning-dependent spectral line, the variation in  $\Gamma_h$ is unexpectedly drastic, varying by two orders of magnitude as the pump field is detuned beyond the $^{14}$N hyperfine splitting frequency (2.16 MHz), while $\Gamma_h$ remains an order of magnitude narrower than $\Gamma\sim6.8$ MHz [Fig. 1(d)].
	
	This measured dependence provides insight into the broadening mechanisms of the ensemble. Considering only a normal
	distribution of frequencies, no drastic change in $\Gamma_h$ should be expected, as usually observed for atomic ensembles. However, when considering the magnetic anisotropy of the N-$V^-$ defect, and the strong dependence of the coherence time on the magnitude of off-axis magnetic fields \cite{Stanwix2010, Tetienne2012}, a detuning-dependent $\Gamma_h$ is not surprising. A possible explanation for this trend is therefore given considering the dominant non-parallel arrangement of the N-$V^-$ symmetry axis and the anisotropic symmetry axis of the P1 defect caused by the Jahn-Teller effect \cite{DeLange2012}. Such miss-alignment is expected to be predominant for N-$V^-$ defects with the closest proximity to P1 defects. These N-$V^-$ defects therefore experience a stronger off-axis magnetic field, simultaneously resulting in a larger Zeeman shift and severer degradation of their coherence time, compared to N-$V^-$ defects that are situated further from their nearest P1 defect.
	
	The observed dependence of $\Gamma_h$ on $(\omega_b-\omega_c)$ is therefore associated with a central portion roughly outlining a frequency range where a large density of transitions have been shifted without significant dephasing, and transition frequencies detuned beyond approximately 3 MHz window, which possess lower $T_2$ times. Considering the disparity between $\Gamma$ and the $\Gamma_h$ trend plotted in Fig. 2(c), it may be considered that the total broadening is dominated by pure dephasing (i.e., a collective degradation of $T_2^*$) introduced by the P1 ensemble \cite{Rosenzweig2018}, rather then just a normal distribution of transitions with similar $T_2$ times.
		
	\section{Continuous Population Oscillations}
	
	\begin{figure*}
		\centering
		\includegraphics[scale = 0.5]{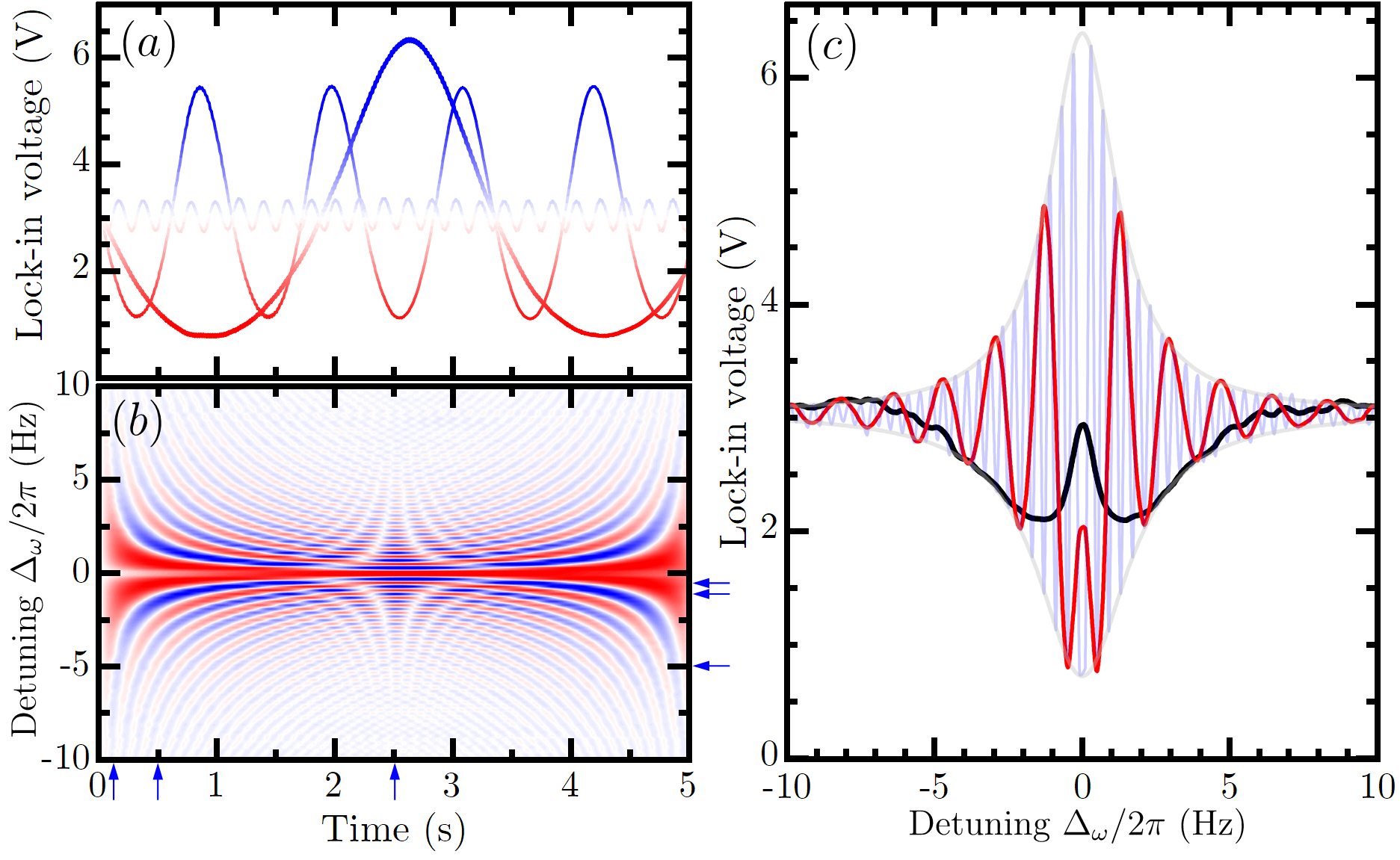}\includegraphics[scale = 0.5]{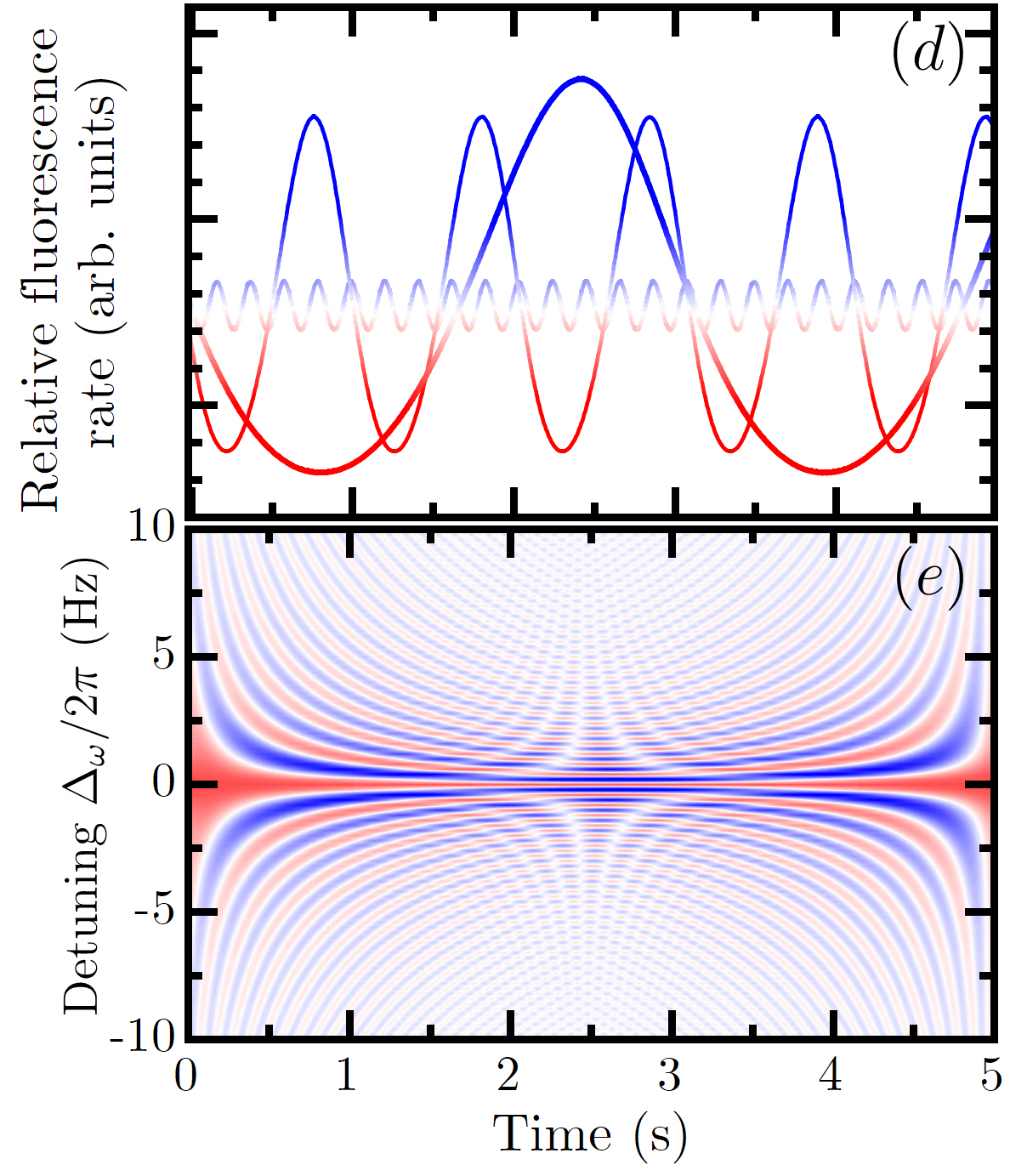}
		\caption{(a) Detected POs of the $\vert0\rangle {\leftrightarrow} \vert {+}1\rangle$ spin transition at the beat frequency $\Delta_\omega$, within the central 'spike' visible in Fig. 2(b). The \textit{y}-scale colour is the same as the \textit{z} colour scale used in (b), which shows the $\Delta_\omega$-time map of the detected POs, and the blue arrows showing the plotted sections in (a) and (c). (c) Phase-continuous plots from sections in (b) annotated by the blue arrows, as a function of $\Delta_\omega$. (d),(e) Simulated $\Delta_\omega$-dependent fluorescence, as detailed in the Appendix, with the set parameters $T_2^* = 40~\mu$s, $\Omega_b/2\pi = \Omega_p/2\pi = 50$ Hz, and $\Lambda/2\pi = 4$ Hz.}  
	\end{figure*}  
	\begin{figure}
		\centering
		\includegraphics[scale = 0.67]{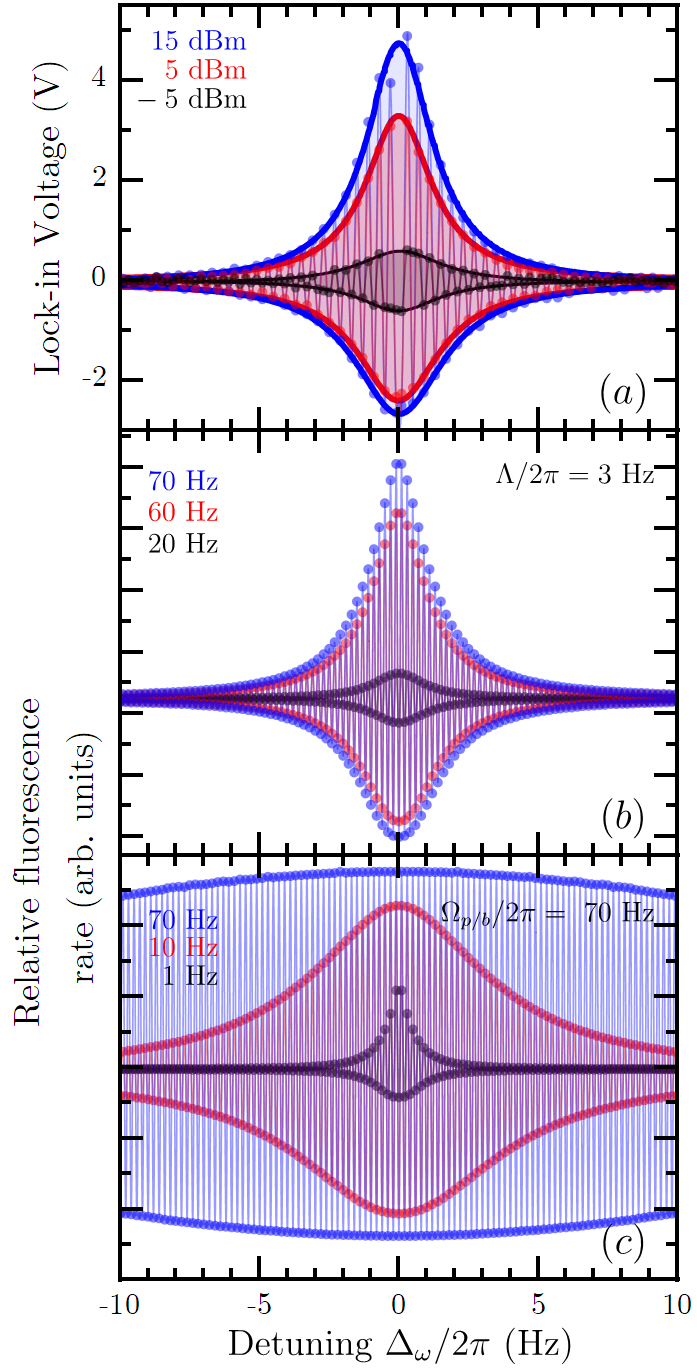}
		\caption{(\textit{a}) Measured $\Delta_\omega$-dependent PO, for three logarithmically increased pump and probe field powers, compared to three simulated traces in (b) with linearly increased $\Omega_{b/p}$. (c) The dependence of the envelope shape on the optical excitation rate $\Lambda$. The simulated traces in (b) and (c) uses $T_2^* = 40~\mu$s.}
	\end{figure}  

	At the centre of each dip in Fig. 2(b) is a sharp spike representing a region where POs are detected. An example is plotted in Fig. 3, which was measured by triggering data accumulation on a set phase of the fluorescence oscillation, enabling the accumulation of in-phase oscillations at $t = 0$, and the mapping of phase-continuous traces as a function of $\Delta_\omega$.
	
	For an isolated two-level system, the ground and exited state occupation should adiabatically vary as the pump and probe fields shift into and out of phase at their beat frequency $\Delta_\omega$. The ground and excited state occupation will then equalise (thermalise) at a rate set by $T_2$, beyond which they remain invariant. In the case of the N-$V^-$ level structure, the presence of a shelving state and its spin state-specific optical transition rates ensures that the system can be continuously polarised into its ground state under continuous incoherent laser excitation \cite{Doherty2013}. This allows for the continuous measurement of POs, as shown in Fig. 3(a), which here resemble rectified sine waves as $\Delta_\omega \rightarrow 0$ Hz. When extracting plots along the $\Delta_\omega$ dimension for a fixed time, a time-dependent density of fringes is observed within a constant envelope. As discussed in Ref. \cite{Mrozek2016}, the shape and features of these oscillations across both dimensions (time and $\Delta_\omega$) is set by both the measurement parameters and the intrinsic coherence times for the subensemble targeted by the probe field. 
	
	\begin{figure*}
		\centering
		\includegraphics[scale = 0.8]{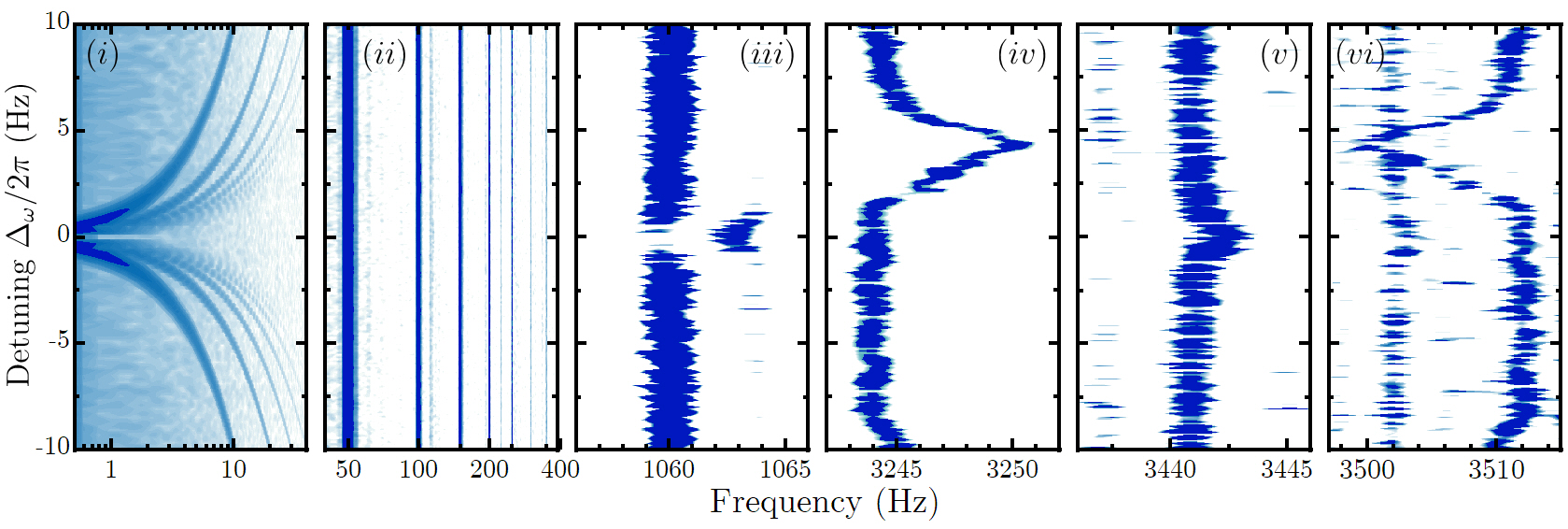}
		\caption{Fourier spectrum maps at various frequency intervals as a function of detuning $\Delta_\omega$. These were obtained using a 10 kHz sampling rate from the directly detected light (no modulation of the probe field, and bypassing the lock-in amplifier).}
	\end{figure*}
	
	These features and their relationships may be understood by attempting to reproduce the trends using a typical five-level model of the N-$V^-$ system \cite{Robledo2011,Jensen2013,El-Ella2017,Ahmadi2017}, while accounting for two coherent drives for the single ground state spin transition:
	\begin{align}
	\hat{\mathcal{H}}/\hbar &= \sum^5_{i=1}\omega_i\vert i\rangle \langle i\vert - D(t)\big(\vert 1\rangle \langle 2\vert + \vert 2\rangle \langle 1\vert\big),\\
	D(t) &= \Omega_p\cos{(\omega_p t)}+\Omega_b\cos{(\omega_b t)}.
	\end{align}
	The construction of this model and the coupled rate equations are outlined in the Appendix, and are summarised here as consisting of two ground-state spin levels, $\vert1\rangle$ and $\vert2\rangle$ that are coherently driven by a microwave field with a Rabi frequency $\Omega_{p/b}$, and two excited-state spin levels $\vert3\rangle$ and $\vert4\rangle$ that are populated with an incoherent optical field at a rate $\Lambda$. The excited state population decays to the ground states both directly and via an intermediate state $\vert5\rangle$ which predominantly accepts from $\vert4\rangle$ and populates $\vert1\rangle$. This model is fitted to the measured data by varying the lifetimes $T_1$ and $T_2^*$, and the excitation rates $\Omega_{p/b}$ and $\Lambda$. Using the experimentally determined $T_2 \sim 40~\mu$s, these measurements were only reproducible using this model by setting $\Lambda$ and $\Omega_{p/b} < 100$ Hz (as
	opposed to the estimated delivered excitation rates in the order of MHz), and by neglecting the longitudinal decay rate such $T_1\gg1$ s was necessary.

	Although this model represents the dynamics of a single N-$V^-$ spin, it successfully reproduces the measured spectrum
	of homogeneous low density N-$V^-$ ensembles \cite{Ahmadi2017, El-Ella2017}. In the case of inhomogeneity and the presence of significant impurity coupling, evident by the measurable P1 electron spin resonances in Fig. 1(b), the model is naturally incomplete. However, the parameters required for fitting this model leads to some insight on the systems dynamics. Primarily, the restriction to low excitation rates accounts for a large absorbing defect density - the required excitation rates for this model may therefore be considered as \textit{effective} excitation rates for single N-$V^-$s in the ensemble, due to the presence of a large density of both N-$V^-$s and non-N-$V^-$ defects that absorb much of the delivered power within the excitation volume.
	
	These effective excitation rates are orders of magnitude slower than the intrinsic level transition rates, and the measured dynamics are therefore by definition adiabatic. Without the incoherent optical excitation and the N-$V^-$ shelving state $\vert5\rangle$, the POs will decay within the $T_2$ dephasing time, and continuous optical repolarization of the ground state is therefore necessary. This may explain the need to exclude $T_1$ in this model when reproducing the measured data, as the continuous dynamic of polarising the ground-state negates the effect of longitudinal decay. Therefore in an adiabatic CW regime, no discernible longitudinal relaxation may be considered to occur, rendering $T_1$ as being effectively infinite ($1/T_1\rightarrow0$). However, this should not be considered as a statement on the inherent $T_1$ of the subensemble, but rather on the incompleteness of the model. Instead, this suggests the importance of the inherent disordered intra-ensemble interactions which have been shown to influence an ensembles collectively measured $T_1$ time - specifically, high disorder has been found to significantly slow down thermalisation and extend $T_1$ as demonstrated in Ref.\cite{Kucsko2018}, using pulsed double resonance driving.
	
	Other factors regulating the measured envelope decay and its asymmetry, include the relation between the optical and spin driving excitation rates with respect to $T_2^*$, as demonstrated in Fig. 4. The width of the envelope is primarily
	modified by the optical excitation rate $\Lambda$, while the amplitude and asymmetry is primarily modified by the Rabi frequencies, provided that these exceed $\Lambda$. Unlike the measurements presented in Ref.\cite{Mrozek2016} where the onset of POs are observed at beat-frequencies in the order of kHz, the measured PO here is observed only at much slower beat frequencies within 20 Hz. Based on the model fitting parameters, this is deduced to be partly a result of the low effective excitation rates due to a large density of non-N-$V^-$ radiation-absorbing centres. This is particularly apparent with respect to the comparative Rabi frequency scaling; the experimental data is taken at 10 dB intervals, whereas to simulate approximately similar shapes and decays, a linear increase in the set Rabi frequency is sufficient.

	\section{Fourier Analysis}
	
	A typical Fourier map of the time-dependent PO is shown in Fig.5, which was measured without modulating the probe field and bypassing the lock-in amplifier, using a 10 kHz sampling rate. The rectified shape of the oscillations leads to the presence of multiple harmonics associated with the beat-note frequency peak, highlighted in Fig.5(i). Another set of directly identifiable peaks are related to the external magnetic “hum” originating from electric mains transformers. This occurs at 50 Hz and is accompanied by multiple harmonics, as highlighted in Fig.5(ii). Being a consistent source associated with all mains-supplied electronics, these frequency components are independent of $\Delta_\omega$, and show no fluctuations beyond their $\leqslant3$ Hz linewidths.
	
	\begin{figure}
		\centering
		\includegraphics[scale = 0.94]{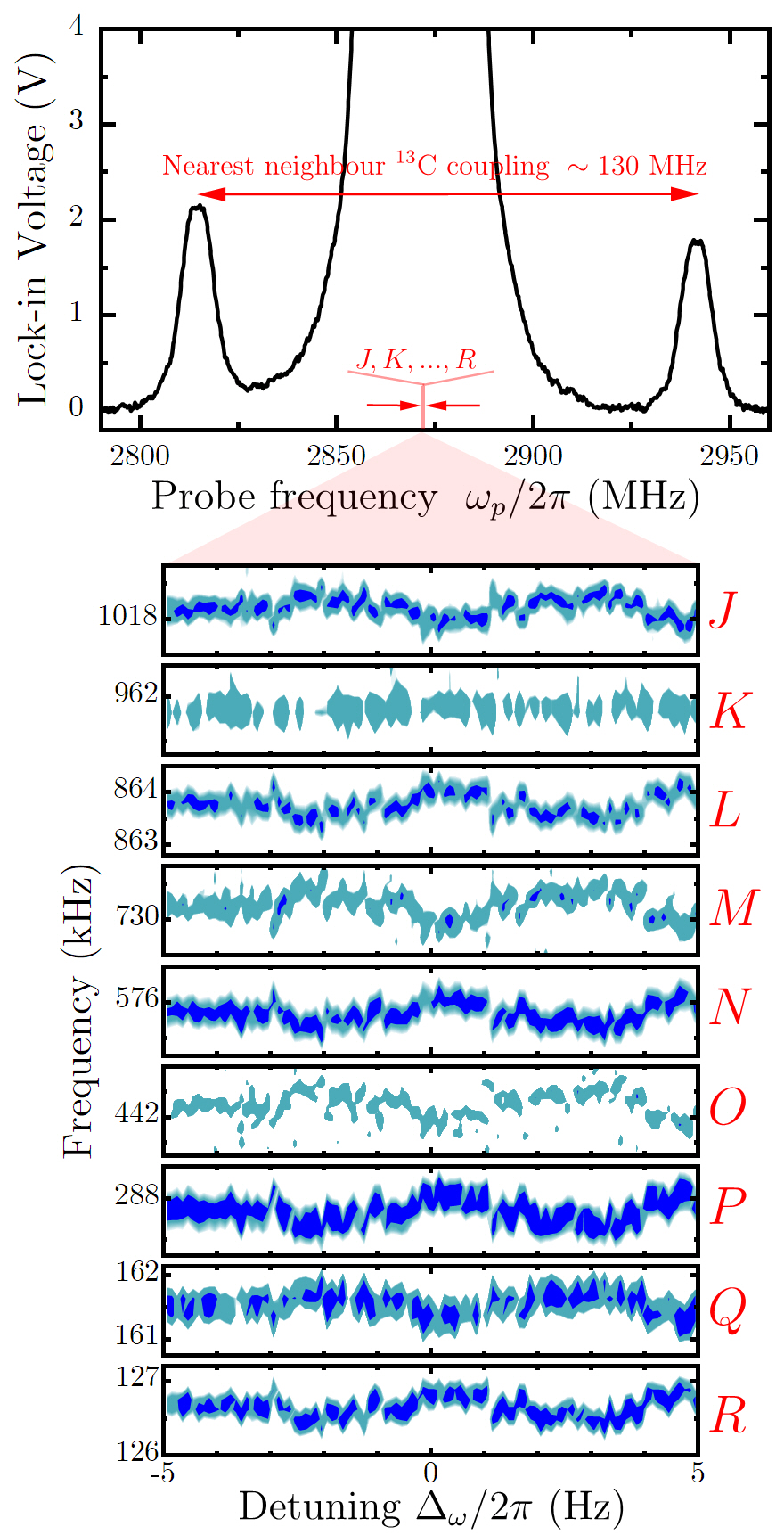}
		\caption{Near zero-field ODMR spectrum highlighting the satellite resonances associated with the hyperfine coupling of $^{13}$C nuclear spins in the nearest-neighbour lattice position to the N-$V^-$ centres. Below are the Fourier spectrum maps of the identified frequency components in the measured PO, designated to $^{13}$C nuclear spins at increasing distances, using the same letter designation as in Ref.\cite{Dreau2012}.}
	\end{figure}  
	
	Much more intriguing is the observed presence of a multitude of discrete frequencies, displaying a range of distinct detuning-dependent behaviour. Identifying these components is carried out by a combination of \textit{a priori} knowledge, measuring their dependence on $\Delta_omega$, and identifying replica peaks which display identical or opposite fluctuations. Two discriminating examples are highlighted in Figs. 5(iii)–5(vi); the symmetrical peaks in Figs. 5(iii) and 5(v) display a shift as$\Delta_\omega \rightarrow 0$, as opposed to an example highlighted in Figs. 5(iv) and 5(vi), which shows no dependence on $\Delta_\omega$, but are replicated at higher frequencies and display symmetric fluctuations. The former are likely dressed-state frequencies mixed with the POs; their frequencies are set by the pump-field Rabi frequency and detuning to the bare transition. The dressed state frequency is expected to shift as the probe and pump fields are tuned into resonance and deliver a larger cumulative Rabi frequency. Here, the shift is measured to be in the order of a few Hz, which corresponds well with the expected dressed state frequency considering the fitted effective Rabi frequencies in the previous section. The latter frequency components in Figs. 5(iv) and 5(vi) are most likely related to a coupled subset of the ensemble with nearby magnetic spins, as they display symmetric fluctuations that is independent of $\Delta_\omega$. These fluctuations are likely due to external magnetic field fluctuations.
	
	A set of directly identifiable frequency components displaying such behaviour can be related to “families” of hyperfine
	coupled $^{13}$C nuclear spins. These are plotted in Fig. 6 and were measured using a 2-MHz sampling rate. These families
	are related to unique separation distances within the crystal lattice and are listed in Table I, which were primarily identified based on the measurements reported by Dr\'{e}au \textit{et al}. \cite{Dreau2012}. The presence of $^{13}$C nuclear spins in the lattice site adjacent the N-$V^-$ vacancy is immediately apparent from the zero-field ODMR measurement, highlighting satellite resonances that are split by the hyperfine interaction energy at approximately 130 MHz. Larger separation distances, with lower hyperfine coupling energies that lie well within $\Gamma$, are observed as discrete frequency components in the PO’s Fourier spectrum and are directly distinguishable due to their symmetric fluctuations, and their independence of $\Delta_\omega$.
	
	\begin{table}
		\caption{\label{tab:table1}%
			Summary of the average measured frequencies (in MHz) compared to those reported in Ref. \cite{Dreau2012}. Numbers in parenthesis indicate the standard deviation of the last digit.}
			\begin{tabular}{c@{\qquad} r@{\qquad}r}
				\toprule
				\textrm{}&
				\textrm{Measured}&
				\textrm{Dr\'{e}au \textit{et al}}\\
				\textrm{$^{13}$C Family}&
				\textrm{@ ${\sim}30$ mT}&
				\textrm{@ 2 mT}\\
				\colrule
				\color{red}{\textit{J}}& ${-}1.0183(4)$ & ${-}1.03(3)$\\
				\color{red}{\textit{K}}& 0.9619(1) & 0.95(2)\\
				\color{red}{\textit{L}}& 0.8638(3) & 0.85(1)\\
				\color{red}{\textit{M}}& ${-}0.7304(4)$ & $-0.70(3)$\\
				\color{red}{\textit{N}}& 0.5758(3) & 0.56(3)\\
				\color{red}{\textit{O}}& ${-}0.4423(5)$ & 0.43(2)\\
				\color{red}{\textit{P}}& 0.2879(4) & $\sim~~~$\\
				\color{red}{\textit{Q}}& ${-}0.1613(2)$ & $\sim~~~$\\
				\color{red}{\textit{R}}& 0.1266(3) & $\sim~~~$\\
				\botrule
			\end{tabular}
	\end{table}

	In particular, the opposite symmetry of the measured fluctuations indicate the sign of the hyperfine coupling constant, representing lattice site positions where the electron density is either negative or positive, as identified through high-resolution pulsed ODMR in Ref. \cite{Dreau2012}. The discrepancy in the magnetic fields used here and reported in Ref. \cite{Dreau2012} may be explained considering the anisotropic coupling of the N-$V^-$ electron and $^{13}$C nuclear spins, and the off-axis magnetic field applied in the measurements reported here.
	
	Curiously, a number of frequency triplets are also observed such as that plotted in Fig. 7, which may be interpreted as a
	signature of a 3/2 spin-coupled system. This is likely related to hyperfine coupling of the $^7$Li nuclear spin, which is an abundant byproduct of the neutron irradiation process. While the resulting hyperfine interaction energy, which in this case is in the order of 16 kHz, depends on the exact spatial separation of the nuclear spin and the N-$V^-$, the resulting peak triplet separation and their relative amplitudes are distinct, and may be attributed to the nuclear spin’s quadrupole moment.
	
	Nuclear spins with spin number $I > 1/2$ possess nonspherical nuclear charge distributions, and therefore a quadrupole moment which results in a hyperfine splitting $\Delta\nu$ that is strongly dependent on the local electric field gradient. The resulting splitting provides insight into the nuclear defects’ immediate electronic environment \cite{Shore1996} and is approximately
	\begin{align}
	\Delta\nu\approx\frac{3eQV_{zz}}{h2I(2I-1)},
	\end{align}
	
	where $Q\sim4\times10^{-30}$ m$^{-2}$ is the quadrupole moment of the $^7$Li nucleus \cite{Stone2005}, $e$ is the elementary charge, $h$ is Plank's constant, and $V_{zz}$ is the principal component of the electric field gradient tensor at the nuclear site \cite{Shore1996}.  By definition, $\Delta\nu$ will only be non-zero when a quadrupole nuclei resides within a non-spherical or non-cubic (not $T_{d}$ or $O_{h}$) symmetric site, and would therefore approach zero at substitutional sites of the diamond lattice \cite{Autschbach2010}. The measured splitting implies an electric field gradient of the order $V_{zz}\sim10^{18}$ Vm$^{-2}$, which reflects interstitial regions approaching the diamond lattice points, as expected due to the small size and high mobility of the $^7$Li atom with respect to the carbon lattice \cite{Stewart1973,Restle1995, Othman2014}. Furthermore, due to second order quadrupole effects, the slight-asymmetry observed in the splitting is expected \cite{Autschbach2010}.
	
	\begin{figure}
		\centering
		\includegraphics[scale = 0.5]{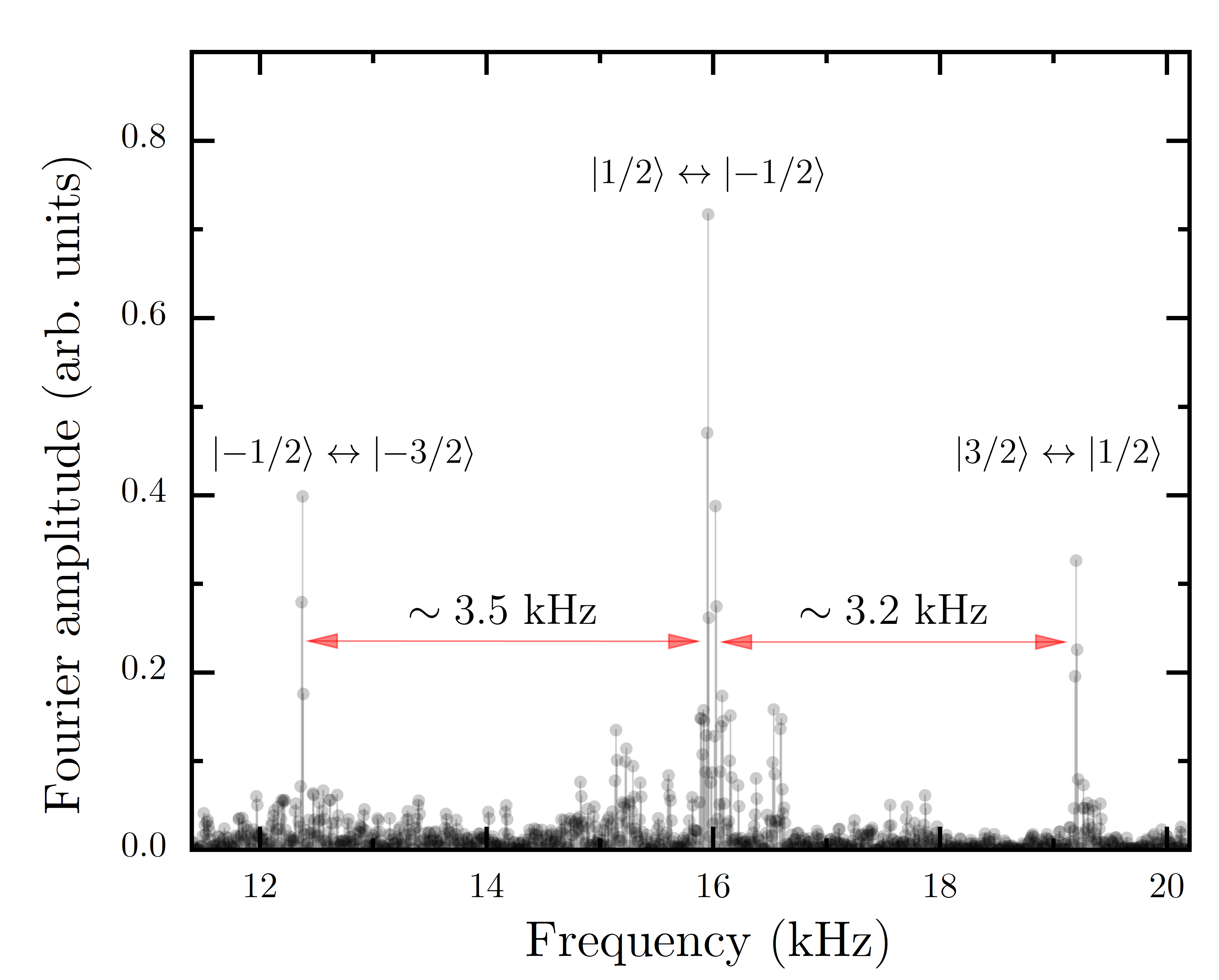}
		\caption{Possible signature of coupled 3/2 nuclear spins, speculated to be related to the neutron irradiation byproduct $^{7}$Li. The asymmetry is a consequence of the second-order quadrupole splitting, as observed in conventional nuclear magnetic resonance \cite{Shore1996}.}
	\end{figure}  
	
	\section{Conclusion}
	The continuous hole-burning experiments presented here have highlighted the detuning-dependent homogeneous linewidth component of an inhomogeneously broadened spin ensemble. This was measured to span from $\sim4$ kHz to $\sim200$ kHz ($T_2\sim40~\mu$s to $\sim0.7~\mu$s) within a 4 MHz detuning of the pump and probe fields, and within a total inhomogeneous linewidth of $\sim7$ MHz ($T_2\sim50$ ns). In particular, the detuning dependence of the probed subensemble’s coherence points toward the role of the magnetic anisotropy of the N-$V^-$ and P1 defects, and the increasingly degraded coherence accompanying larger Zeeman shifts for N-$V^-$s situated closer to P1 defects.
	
	Under continuous laser excitation, adiabatically driven POs were observed to be unresolvable beyond a 10 Hz detuning
	of the pump and probe fields. These measurements were analysed by numerically reproducing them using a five-level
	system of coupled rate equations. The necessary optical and microwave excitation rates needed to obtain a good fit with
	the observed data were in the order of 10 Hz, in contrast to the experimentally delivered powers with estimated excitation rates in the order of MHz. This points toward the possibility that a very low effective excitation rate of the ensemble subset was achieved, in light of the confirmed dominant presence of absorbing emitters other than N-$V^-$s. Furthermore, the need to exclude the longitudinal relaxation rate $T_1$ to reproduce the experimental data highlighted the limitation of the model in describing the adiabatic dynamics for this particular system. This was considered to be due to the model’s incompleteness in accounting for the disordered intraensemble interactions, which has been previously reported to influence an ensembles thermalisation, and collective $T_1$ time. Fourier analysis of these oscillations highlighted a range of unique spectral components of which some were directly identified to stem from $^{13}$C nuclear hyperfine coupling, and others speculated to arise from the presence of $^7$Li, a byproduct of neutron irradiation. 
	
	Ultimately, this paper demonstrates the complimentary utility of continuous CW spectral hole burning to more common pulsed double-resonance schemes, for studying inhomogeneous ensembles and their sublinewidth coupled states,
	while the demonstration of a detuning dependence points towards possible practical magnetic field-based metrological use. Further experimental and theoretical work will be carried out to develop a more complete description of the ensemble system investigated here and its adiabatic dynamics, and to more rigorously identify all spectral components of the POs up to bandwidths approaching the $^{14}$N hyperfine coupling frequency.

	\begin{acknowledgments}
		This work was funded by the Villum Foundation (Grant No. 17524). U.L.A. would like to acknowledge funding from the Danish Research Council through a Sapere Aude grant (DIMS, Grant No. 4181-00505B), and the Danish National Research Foundation (bigQ, DNRF142). We are grateful to Kristian Hagsted Rasmussen for sample preparation and to Ilja Radko and Rasmus Jensen for helpful discussions.
	\end{acknowledgments}
\appendix*
\section{N-$V^-$ coupled rate equations}
\begin{figure}[h!]
	\centering
	\includegraphics[scale = 0.45]{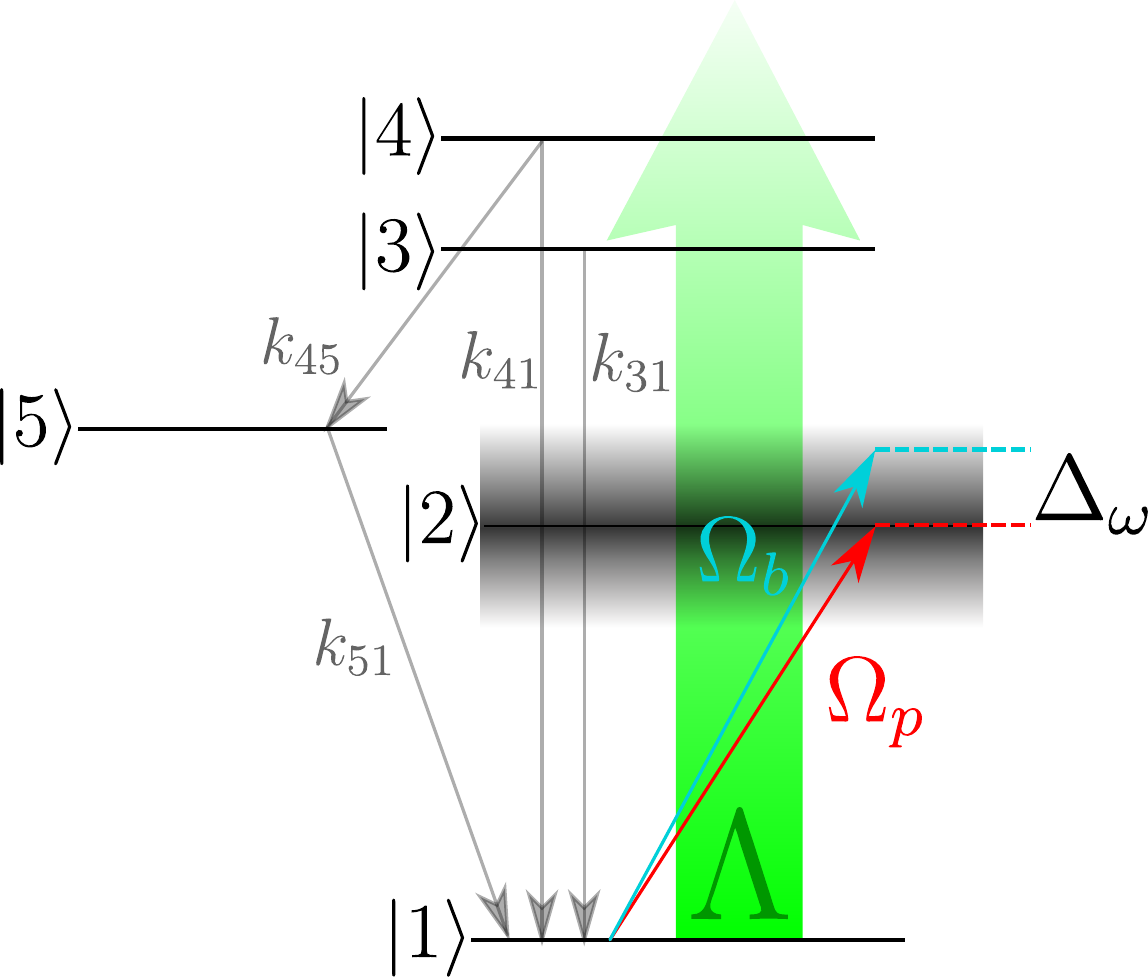}
	\caption{Depiction of the five-level model described by the Hamiltonian in Eq. (2) and used to construct the coupled rate equations (only some of the rates are annotated).}
\end{figure}  	
	The Hamiltonian defined in Eq. (2), and its set of coupled equations of motion is derived according to the level arrangement illustrated in Fig. 8, considering two in-phase sinusoidal drives simultaneously applied as described in Eq. (3). This model consists of two coherently driven ground-state spin levels $\vert 1\rangle \leftrightarrow \vert 2\rangle$, and two excited spin state levels $\vert 3\rangle$ and $\vert 4\rangle$ which are populated incoherently using an optical excitation rate. These decay both directly to $\vert 1\rangle$ and $\vert 2\rangle$, and also via the shelving state $\vert 5\rangle$ which decays to both ground states, but predominantly from $\vert 4\rangle$ and towards $\vert 1\rangle$.

	The evolution of a given density matrix $\hat{\rho}$ is then derived in terms of Heisenberg-Langevin equations of motion, and in the rotating frame of the probe field $\omega_p$. Applying the rotating wave approximation, the coupled set of equations are derived as
\small \begin{align}
	\dot{\rho}_{11} &= -\Lambda\rho_{11}+k_{31}\rho_{33}+k_{41}\rho_{44}+k_{51}\rho_{55}\nonumber\\
	&-\frac{\pi}{T_1}(\rho_{11}-\rho_{22})-\frac{i}{2}\rho_{12}'(\Omega_p + \Omega_be^{i\Delta_\omega t})\\
	&+\frac{i}{2}\rho_{21}'(\Omega_p+\Omega_be^{-i\Delta_\omega t}),\nonumber\\
	\dot{\rho}_{22} &= -\Lambda\rho_{22}+k_{32}\rho_{33}+k_{42}\rho_{44}+k_{52}\rho_{55}\nonumber\\
	&-\frac{\pi}{T_1}(\rho_{22}-\rho_{11})+\frac{i}{2}\rho_{12}'(\Omega_p + \Omega_be^{i\Delta_\omega t})\\
	&-\frac{i}{2}\rho_{21}'(\Omega_p+\Omega_be^{-i\Delta_\omega t}),\nonumber\\
	\dot{\rho}_{12} &= -(\gamma_2'-i\delta_c)\rho_{12} +\frac{i}{2}(\rho_{22}-\rho_{11})(\Omega_p +\Omega_be^{i\Delta_\omega t}),
	\\
	\dot{\rho}_{21} &= -(\gamma_2' + i\delta_c)\rho_{21}  -\frac{i}{2}(\rho_{22}-\rho_{11})(\Omega_p +\Omega_be^{-i\Delta_\omega t}),
	\\
	\dot{\rho}_{33} &= \Lambda\rho_{11}-(k_{35}+k_{32}+k_{31})\rho_{33},
	\\
	\dot{\rho}_{44} &= \Lambda\rho_{22}-(k_{45}+k_{42}+k_{41})\rho_{44},
	\\
	\dot{\rho}_{55} &= k_{45}\rho_{44}+k_{35}\rho_{33} - (k_{52}+k_{51})\rho_{55},
\end{align}
where $\delta_c = \omega_p-\omega_c$ is the detuning of the probe beam frequency to the central spin transition frequency, while $\Delta_\omega = \omega_b-\omega_p$ is the detuning between the pump and probe frequencies. The total effective dephasing rate is a sum of the dephasing rate, and the contribution of the optical excitation $\gamma_2' = 2\pi(1/2T_1 + 1/T_2^*) + \Lambda/2$. The time-dependent
fluorescence is obtained from the sum  $\mathcal{I} =\alpha\rho_{33} + \beta\rho_{44}$, where $\alpha$ and $\beta$  are the ratio’s of the given levels direct spin-conserving decay rates to the sum of all its decay rates. Values for the level decay rates $k_{mn}$ are taken from Ref. \cite{Robledo2011}, and the remaining rates where varied to reproduce the experimental measurements, except for $T_1$ which was required to be set $>10$ s to obtain a good fit.
\bibliography{ArxivV2}	
\end{document}